\documentclass[runningheads]{llncs}
\usepackage{graphicx}
\usepackage{amsmath}
\usepackage{algorithm}
\usepackage[noend]{algpseudocode}
\makeatletter
\def\BState{\State\hskip-\ALG@thistlm}
\makeatother

\begin{document}
\title{The KiTS19 Challenge Data:\\ 300 Kidney Tumor Cases with Clinical\\ Context, CT Semantic Segmentations, and Surgical Outcomes}
\titlerunning{The KiTS19 Challenge Data}
\author{
Nicholas Heller\inst{1} \and
Niranjan Sathianathen\inst{1} \and
Arveen Kalapara\inst{1} \and
Edward Walczak\inst{1} \and
Keenan Moore\inst{2} \and
Heather Kaluzniak\inst{3} \and
Joel Rosenberg\inst{1} \and
Paul Blake\inst{1} \and
Zachary Rengel\inst{1} \and
Makinna Oestreich\inst{1} \and
Joshua Dean\inst{1} \and
Michael Tradewell\inst{1} \and
Aneri Shah\inst{1} \and
Resha Tejpaul\inst{1} \and
Zachary Edgerton\inst{1} \and
Matthew Peterson\inst{1} \and
Shaneabbas Raza\inst{3} \and
Subodh Regmi\inst{1} \and
Nikolaos Papanikolopoulos\inst{1} \and
Christopher Weight\inst{1}
}
\authorrunning{N. Heller et al.}
\institute{
University of Minnesota \and
Carleton College \and
University of North Dakota\\
\email{\{helle246, cjweight\}@umn.edu}
}
\maketitle 
\begin{abstract} % 170/170 words

Characterization of the relationship between a kidney tumor's appearance on cross-sectional imaging and it's treatment outcomes is a promising direction for informing treatement decisions and improving patient outcomes. Unfortunately, the rigorous study of tumor morphology is limited by the laborious and noisy process of making manual radiographic measurements. Semantic segmentation of the tumor and surrounding organ offers a precise quantitative description of that morphology, but it too requires significant manual effort. A large publicly available dataset of high-fidelity semantic segmentations along with clinical context and treatment outcomes could accelerate not only the study of how morphology relates to outcomes, but also the development of automatic semantic segmentation systems which could enable such studies on unprecedented scales. We present the KiTS19 challenge dataset: a collection of segmented CT imaging and treatment outcomes for 300 patients treated with partial or radical nephrectomy between 2010 and 2018. 210 of these cases have been released publicly and the remaining 90 remain private for the objective evaluation of prediction systems developed using the public cases. 

\end{abstract}
\section{Background \& Summary} % 452/700 words
\label{background}
% Info about kidney cancer, nephrometry, existing biomarkers

There were more than 400,000 kidney cancer diagnoses worldwide in 2018 resulting in more than 175,000 deaths \cite{bray2018global}, up from 208,000 diagnoses and 102,000 deaths in 2002 \cite{parkin2005global}. The incidence is higher in developed countries than in developing countries, and peaks between the ages of 60 and 70 \cite{capitanio2016renal}. With the increase in abdominal imaging for various unrelated indications, the incidental detection of asymptomatic renal masses has become increasingly common. This has increased the proportion of tumors that are small and localized when treated, which is thought to be a contributing factor to the disease's increased overall survival \cite{homma1995increased}. Some established risk factors for kidney cancer are smoking, obesity, and hypertension \cite{chow2010epidemiology}. Historically, removal of both the tumor and affected kidney, termed Radical Nephrectomy (RN), was standard of care for kidney tumors, but advancements in surgery in conjunction with earlier tumor detection have precipitated shift in kidney cancer treatment toward more conservative nephron sparing procedures, termed Partial Nephrectomies (PNs) \cite{sun2012treatment}. These are typically less invasive and limit renal function impairment, thus they are preferred when feasible.

In an effort to more reliably quantify tumor details and accurately compare decisions about kidney tumor treatment (notably, the decision between RN and PN), various \textit{nephrometry} scoring systems were proposed based on the tumor's presentation in cross-sectional imaging. Among these are R.E.N.A.L. \cite{renal}, P.A.D.U.A. \cite{padua}, and the Centrality Index \cite{cindex}. Once proposed, these scoring systems were found to be associated with surgical approach and a wide range of clinical outcomes including recurrence after treatment \cite{maxwell2016renal,gahan2015performance}, benign vs malignant tumor \cite{osawa2016comparison}, and high-grade tumor pathology \cite{kutikov2011anatomic}. In spite of their impressive predictive power, existing nephrometry scores characteristically utilize relatively simple and easy-to-extract image features such as location, degree of endophycity, and diameter of the tumor. Further, the most popular R.E.N.A.L. and P.A.D.U.A. scores reduce continuous variables into discrete bins, further limiting their expressive power in favor of more expedient and repeatable manual evaluation \cite{joshi2017renal}. In contrast, semantic segmentation produces a rich quantitative representation of the tumor and affected organ, trivializing the precise extraction of enumerable morphological traits such as contact surface area \cite{leslie2014renal} and irregularity \cite{yap2018quantitative}. 

Our objectives in releasing this data are (1) to accelerate the research and development of new nephrometric features to aid in prognosis and treatment planning for kidney tumors, and (2) to enable the creation of reliable learning-based kidney and kidney tumor semantic segmentation methods which will allow the features developed in (1) to be automated and applied at an unprecedented scale. 

\section{Methods} % no limit
\label{methods}
% Inclusion criteria, imaging collection/annotation methods, chart review
We conducted a retrospective review\footnote{This work was reviewed and approved by the Institutional Review Board at the University of Minnesota as Study 1611M00821.} of the 544 patients who underwent RN or PN at our institution between 2010 and mid-2018 to excise a renal tumor. For 326 of these patients, preoperative abdominal CT imaging in the late-arterial phase was available, and the remaining patients were excluded. To simplify an unambiguous definition of kidney tumor voxels, all patients with tumor thrombus were also excluded, leaving 300 patients who met our inclusion criteria and comprise our dataset.

The data collection procedure for each included patient consisted of four steps: (1) chart review, (2) CT collection, (3) CT annotation, and (4) quality assurance. This work was done primarily by medical students under the supervision of author Christopher Weight, an experienced fellowship-trained urologic oncologist who specializes in kidney tumors.

\subsection{Chart Review}
\label{methods__chart_review}
The objective of the chart review phase of the data collection procedure was to record relevant clinical information about each patient's demographics, comorbidities, intervention, and clinical outcomes. This information was found by a manual review of each patient's Electronic Medical Record (EMR) in conjunction with a database query for certain structured fields. An exhaustive list and short description of each of the collected attributes from this phase can be found in Section \ref{data_records}.

\subsection{CT Collection}
\label{methods__ct_collection}
The objective of the CT collection phase of the data collection procedure was to secure a local copy of the most recent preoperative CT study for each patient that contained at least one series in late arterial contrast phase that depicts the entirety of the abdomen (at least). Despite the fact that such imaging is standard of care for kidney tumors \cite{capitanio2016renal}, many patients were excluded at this stage because it was either done at a referring institution and not available to our team, or MRI was used instead for preoperative planning. 

In rare cases where several preoperative studies captured within one week of each other meet this criteria, preference was given to the study containing the late arterial series with smallest slice-thickness.

\subsection{CT Annotation}
\label{methods__ct_annotation}
Once a patient's clinical attributes and imaging were collected, they were moved to the third phase of our data collection procedure: manual delineation of the kidney and tumor boundaries. To perform these annotations in a distributed manner, we developed a simple web application based on the HTML5 \textit{Canvas} element that allowed users to draw freehand contours on images \cite{heller2017web}. All annotations were performed in the transverse plane, and series were regularly subsampled in the longitudinal direction such that the number of annotated slices depicting any kidney was roughly 50 per patient. Interpolation (described later) was performed to compute labels for the excluded slices.

\subsubsection{Manual Delineation}
\label{methods__ct_annotation__manual_delineation}
\begin{figure}
\includegraphics[width=\textwidth]{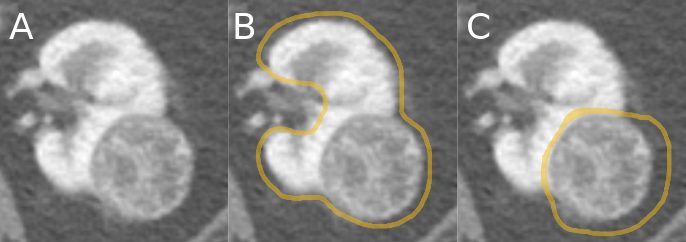}
\caption{\textbf{Left:} An axial section of a kidney and tumor from the database, $I^{(i)}_j$. \textbf{Middle:} An example of the first contour the students were instructed to draw around the whole renal capsule and tumor but excluding the intra-hilar structures, $\partial A^{(i)}_{j,0}$. \textbf{Right:} An example of the second contour the students were instructed to draw which includes the tumor but excludes all kidney tissue, $\partial C^{(i)}_{j,0}$.} \label{fig:kidney_and_tumor_manual}
\end{figure}

The students performing these annotations were given the following instructions:

\begin{enumerate}
    \item Confirm that the collection of images for this patient depicts the entirety of all kidneys. Some of the patients found in our review had horseshoe kidneys or were transplant recipients. These were included so long as the entirety of all kidneys were shown. Let the $i$th case have $J$ transverse slices. We will refer to the voxels from the $j$th slice of the $i$th case as $I^{(i)}_j$.
    \item For each connected component of pixels belonging to a region of interest, draw a contour which \textit{includes} the entire renal capsule and any renal tumors or cysts, but \textit{excludes} all tissue other than renal parenchyma that appears more radiodense than the perinephric fat. In slices where the hilum was present, the students were to introduce a concavity so as to exclude the bright ureter and renal vessels (see Fig. \ref{fig:kidney_and_tumor_manual}b). Let there be $N^{(i)}_{j}$ such contours in the $j$th axial section of the $i$th case. We will refer to the set of voxels inside one of these contours by $A^{(i)}_{j,n}$.
    \item For each connected component of tumor tissue, draw a contour which \textit{includes} that tumor component, but \textit{excludes} all kidney tissue. Effectively, these contours only specify the interface between the kidney and tumor, since the rest of the tumor boundary was already specified in step 2 (see Fig. \ref{fig:kidney_and_tumor_manual}c). Let there be $M^{(i)}_{j}$ such contours in the $j$th image of the $i$th case. We will refer to the set of voxels inside one of these contours by $C^{(i)}_{j,m}$.
\end{enumerate}

This annotation procedure enabled the students to provide a complete and unambiguous representation of the kidneys and kidney-tumor boundary while limiting the number of tedious, voxel-wise decisions. 

\subsubsection{Thresholding and Hilum Filling}
\label{methods__ct_annotation__thresholding_and_hilum_filling}
It is well-established that kidneys, tumors, and cysts (HU \textgreater$\,$ 0.0) are significantly more radiodense than fat (HU \textless$\,$ -90.0) \cite{lepor2000prostatic}, thus a simple HU threshold can be used to precisely define the boundary between the two. Certain CT series, those especially captured with low-dose techniques \cite{lu2001noise}, exhibit random noise which can degrade the performance of the threshold-defined boundary. To mitigate this, we convolve a 3x3 mean filter with each slice before performing a threshold. Experimentally, we determined that a cutoff of -30.0 HU successfully discriminated perinephric fat and the tissue in our regions of interest. In certain cases where the CT had a large amount of noise and no cysts were present, we applied a 7x7 median filter to each slice and raised the threshold value to 0 HU. We will refer to the set voxels in the $j$th slice of the $i$th case found to be above its respective threshold as $S^{(i)}_{j}$.

Between the manually-drawn contours and the thresholding, we partition the voxels from our annotated slices into three bins:

\begin{enumerate}
    \item Loose Background, $B_{loose}$, a superset of True Background $B$, everything outside of the intersection between the thresholded voxels, $S^{(i)}_j$ and the union of all kidney+tumor contour interiors, $A^{(i)}_j$
    $$B^{(i)}_{j} \subseteq B^{(i)}_{loose,j} = I^{(i)}_{j} \setminus \left(\bigcup_{m=1}^{M^{(i)}_{j}} A^{(i)}_{j,m} \cap S^{(i)}_{j}\right)$$
    \item True Tumor, $T$, the intersection of the tumor contour interiors, $C^{(i)}_j$, with the kidney contour interiors, $A^{(i)}_j$, and threshold, $S^{(i)}_j$
    $$T^{(i)}_{j} = \left(\bigcup_{m=1}^{M^{(i)}_{j}} A^{(i)}_{j,m} \cap S^{(i)}_{j}\right) \bigcap \left( \bigcup_{n=1}^{N^{(i)}_{j}} C^{(i)}_{j,n}\right)$$
    \item Strict Kidney, $K_{strict}$, a subset of true kidney, $K$, voxels which appear in the intersection of the kidney contour interiors and threshold but not the tumor contour interiors.
    $$K^{(i)}_{j} \supseteq K^{(i)}_{strict,j} = \left(\bigcup_{m=1}^{M^{(i)}_{j}} A^{(i)}_{j,m}\cap S^{(i)}_{j}\right)\bigcap \left( \bigcap_{n=1}^{N^{(i)}_{j}} I^{(i)}_{j}\setminus C^{(i)}_{j,n}\right)$$
\end{enumerate}
\vspace{1em}

\begin{figure}
\includegraphics[width=\textwidth]{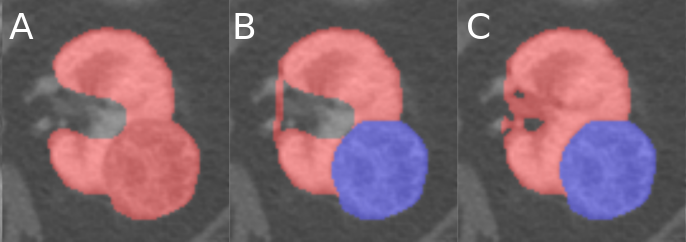}
\caption{A demonstration of the various stages of the algorithm which produces the ground truth segmentation masks given the manually-drawn contours, best viewed in color. \textbf{Left:} The union of all the strict kidney sets, $K^{(i)}_{strict,j}$. \textbf{Middle:} The hilum found by the heuristic based detection algorithm as well as the true tumor found by the intersection of the tumor contour with the left figure, $K^{(i)}_{strict,j} \cup H^{(i)}_{j}$. \textbf{Right:} The final kidney and tumor labels found by including all tissue within the hilum that's above the threshold, blue: $T^{(i)}_{j}$, red: $K^{(i)}_{j}$.} \label{fig:kidney_and_tumor_algo}
\end{figure}

\noindent These bins are depicted in Fig. \ref{fig:kidney_and_tumor_algo}b. Consider the kidney or cyst voxels excluded from $K_{strict}$. We refer to these as $K^{(i)}_{exc,j} = K^{(i)}_{j} \setminus K^{(i)}_{strict,j}$. By definition:
\begin{align*}
K^{(i)}_{j} &= K^{(i)}_{strict,j} \cup K^{(i)}_{exc,j}\\ 
B^{(i)}_{j} &= B^{(i)}_{loose,j} \setminus K^{(i)}_{exc,j}
\end{align*}

\noindent Therefore, if we identify $K_{exc}$, we can compute the final ground truth partition, $B, K, \text{and}\; T$ for each annotated slice. 

On inspection and trial we found that reliably delineating the boundary between the complex intra-hilar structures and kidney parenchyma is not feasible, and to attempt this would only introduce ambiguity and error into our dataset, something that's been shown to markedly hinder the performance of deep learning-based automatic segmentation \cite{heller2018imperfect}.

To address this, we chose to include these intra-hilar structures in our ``kidney'' label. We define the boundary for these features to be that line which spans the concavity formed by the exclusion of this tissue in the manually-drawn contours (see Fig. \ref{fig:kidney_and_tumor_algo}b). This line, $H^{(i)}_{j}$, is computed by a call to OpenCV's \texttt{convexHull()} function followed by \texttt{convexityDefects()}. An heuristic approach based on location and shape was used to automatically select the correct defect and these were manually checked and corrected where necessary. Thus, $K^{(i)}_{j}$ is defined by the inclusive interior of the contour given by $\partial K^{(i)}_{strict,j} \cup H^{(i)}_{j}$, where $\partial K$ denotes the set $K$'s boundary.

\subsubsection{Interpolation}
\label{methods__ct_annotation__interpolation}
Until now, we have described only the procedure defining the ground truth given these manually drawn contours, but for practical reasons only a fraction of the total number of slices containing a region of interest were annotated. In order to produce contours for the remaining slices, and interpolation methodology was used. Our algorithm for interpolating contours for the $l$th slice from contours drawn in slices $l+a$ and $l-b$ in Algorithm \ref{algo:interpolate_contours}. Once these contours are inferred, the ground truth is computed just as it is for manually provided contours.

\begin{algorithm}
\caption{Interpolate Contours}\label{algo:interpolate_contours}
\begin{algorithmic}[1]
\Function{MatchContourPoints}{contour1, contour2}
\State result $\gets$ [] 
\For{$\vec{x}$ in contour1}
\State mindist $\gets inf$
\State $\vec{m} \gets \text{Centroid(contour1)}$
\For{$\vec{y}$ in contour2}
\If{$\left\lVert \vec{x}-\vec{y}\right\lVert_2 < \text{mindist}$}
\State $\text{mindist} \gets \left\lVert \vec{x}-\vec{y}\right\lVert_2$
\State $\vec{m} \gets \vec{y}$
\EndIf
\EndFor
\State Append($(\vec{x},\vec{m})$, result)
\EndFor
\Return result
\EndFunction
\State \# Populates $A^{(i)}_{l}$, $C^{(i)}_{l}$, W.L.O.G Assume $M^{(i)}_{l+a} \geq M^{(i)}_{l+a}$, $N^{(i)}_{l+a} \geq N^{(i)}_{l+a}$
\State \# Distance between contours is taken the euclidean distance between centroids
\State $D_{max} \gets 20$ \# Maximum distance where contours are still morphed together 
\State $A^{(i)}_{l} \gets \{\}$ 
\For{$m$ in $\{1...M^{(i)}_{l+a}\}$}
\State $P \gets$ nearest of contours from $A^{(i)}_{l-b}$ or $\{\}$ if nearest is farther than $D_{max}$
\State $R \gets$ MatchContourPoints($A^{(i)}_{l-b,m}$, $P$)
\State $A^{(i)}_{l} \gets A^{(i)}_{l} \cup \left\{\frac{a}{a+b}*R[:,1] + \frac{b}{a+b}*R[:,2]\right\}$ 
\EndFor
\For{$n$ in $\{1...N^{(i)}_{l+a}\}$}
\State $P \gets$ nearest of contours from $C^{(i)}_{l-b}$ or $\{\}$ if nearest is farther than $D_{max}$
\State $R \gets$ MatchContourPoints($C^{(i)}_{l-b,n}$, $P$)
\State $C^{(i)}_{l} \gets C^{(i)}_{l} \cup \left\{\frac{a}{a+b}*R[:,1] + \frac{b}{a+b}*R[:,2]\right\}$
\EndFor
\end{algorithmic}
\end{algorithm}

\subsection{Code Availability}

Make code available for this hilum filling and interpolation. Have own github, be able to run a demo

\subsection{Quality Assurance}
\label{methods__quality_assurance}

\subsubsection{Chart Review}
\label{methods__quality_assurance__chart_review}
During chart review, the students were instructed to leave blank any field that they were not certain about. These fields were then revisited at a later time by two students. If those students did not agree on the field's correct value, author Christopher Weight was consulted to make the final determination. 

\subsubsection{Imaging Annotations}
\label{methods__quality_assurance__imaging_annotations}
Students performing annotations were instructed to read the radiology note from the preoperative CT scan in order to properly locate and delineate the tumor(s) in concordance with the expert clinician. A reviewing student examined each and every image-ground truth pair in both the transverse and coronal planes, checking for consistent boundary treatment, and once again for concordance with the radiologist's impression. Cases found to have minor issues were fixed by this reviewing student directly, and then accepted, whereas rare cases with major issues were sent back to the first student for fixing, and subsequent re-review. This second practice helped to not only reduce the annotation burden on the reviewing student, but also to educate the annotating students and prevent similar issues in the future. We discuss our method's interobserver variability in section \ref{technical_validation}.

\section{Data Records}
\label{data_records}

The imaging and semantic segmentation labels that were used for the 2019 KiTS Challenge were originally released on GitHub\footnote{https://github.com/neheller/kits19}. The kits19 repository contains a directory named \texttt{data/} which has a subdirectory for each case using the naming convention e.g. \texttt{case\_00123} for case 123. Cases are numbered beginning at 0, and the first 210 cases (\texttt{case\_00000} - \texttt{case\_00209}) comprise the public portion of the dataset. Within each subfolder is the case's imaging and segmentation labels (named \texttt{imaging.nii.gz} and \texttt{segmentation.nii.gz} respectively) as well as a JSON file with that case's clinical attributes. A comprehensive specification of that JSON file can be found below.

\begin{itemize}
    % CHECK 
    \item \texttt{case\_id} (String): A unique identifier for each case. This takes the form of \texttt{"case\_"} followed by five digits, where the least significant digits correspond to the case index and unused digits are assigned zero. For instance, \texttt{"case\_00000"}, \texttt{"case\_00017"}, \texttt{"case\_00202"}

    % CHECK
    \item \texttt{age\_at\_nephrectomy} (Integer): The age of the patient at the time that they underwent nephrectomy for their renal tumor.
    
    % CHECK
    \item \texttt{gender} (Categorical): The gender of the patient. This takes one of the following values: \texttt{\{"male", "female"\}}
    
    % CHECK
    \item \texttt{body\_mass\_index} (Float): The body mass index of the patient at the time measured nearest to the most recent imaging in the dataset.
    
    % CHECK
    \item \texttt{comorbidities} (Object - Bitmap): This takes an object with the following boolean attributes: \texttt{myocardial\_infarction}, \texttt{congestive\_heart\_failure}, \texttt{peripheral\_vascular\_disease}, \texttt{cerebrovascular\_disease}, \texttt{dementia},\\ \texttt{copd}, \texttt{connective\_tissue\_disease}, \texttt{peptic\_ulcer\_disease},\\ \texttt{uncomplicated\_diabetes\_mellitus},\\ \texttt{diabetes\_mellitus\_with\_end\_organ\_damage}, \texttt{chronic\_kidney\_disease},\\ \texttt{hemiplegia\_from\_stroke}, \texttt{leukemia}, \texttt{malignant\_lymphoma},\\ \texttt{localized\_solid\_tumor}, \texttt{metastatic\_solid\_tumor}, \texttt{mild\_liver\_disease}, \texttt{moderate\_to\_severe\_liver\_disease}, \texttt{aids}
    
    % CHECK
    \item \texttt{smoking\_history} (Categorical): This attribute can take any of the following values: \texttt{\{"never\_smoked", "previous\_smoker", "current\_smoker"\}}.
    \item \texttt{age\_when\_quit\_smoking} (Integer): The age at which the patient quit smoking. This takes the value of \texttt{"not\_applicable"} for cases in which it is not applicable and \texttt{null} for cases in which it's not known (22 instances). 
    \item \texttt{pack\_years} (Integer): An estimate of the number of cigarette pack-years that this patient has smoked. This takes the value \texttt{null} if it is unknown (67 instances).
    \item \texttt{chewing\_tobacco\_use} (Categorical): This attribute can take any of the following values: \texttt{\{"never\_or\_not\_for\_more\_than\_3mo",  "quit\_in\_last\_3mo", "currently\_chews"\}}.
    \item \texttt{alcohol\_use} (Categorical): This attribute can take any of the following values: \texttt{\{"never\_or\_not\_in\_last\_3mo", "two\_or\_less\_daily", \\"more\_than\_two\_daily", "quit\_in\_last\_3mo"\}}.
    
    % CHECK
    \item \texttt{intraoperative\_complications} (Object - Bitmap): This takes an object with the following boolean attributes: \texttt{blood\_transfusion},\\ \texttt{injury\_to\_surrounding\_organ}, \texttt{cardiac\_event}
    
    % CHECK
    \item \texttt{hospitalization} (Integer): The number of days this patient spent in the hospital after their nephrectomy operation. If the patient died before being discharged from the hospital, this attribute will take the value\\ \texttt{"died\_before\_discharge"}.
    
    % CHECK
    \item \texttt{ischemia\_time} (Integer): The number of minutes that the kidney was deprived of blood during the nephrectomy operation. This takes the value of \texttt{"not\_applicable"} for radical nephrectomies and \texttt{null} for partial nephrectomies for which this value is not available (10 instances).

    % TODO...
    \item \texttt{radiographic\_size} (Float): The size of the tumor reported in the radiology report.
    \item \texttt{pathologic\_size} (Float): The size of the tumor reported in the surgical pathology report.
    
    % CHECK
    \item \texttt{malignant} (Boolean): \texttt{true} if the post-operative surgical pathology report indicates that the tumor was malignant, \texttt{false} otherwise.

    % CHECK
    \item \texttt{pathology\_t\_stage} (Categorical): The T-stage reported in the post-operative surgical pathology report. This takes one of the following \texttt{\{"X", "0", "1a", "1b", "1c", "2a", "2b", "3", "4"\}}
    \item \texttt{pathology\_n\_stage} (Categorical): The N-stage reported in the post-operative surgical pathology report. This takes one of the following \texttt{\{"X", "0", "1"\}}
    \item \texttt{pathology\_m\_stage} (Categorical): The M-stage reported in the post-operative surgical pathology report. This takes one of the following \texttt{\{"X", "0", "1"\}}
    
    % CHECK
    \item \texttt{tumor\_histologic\_subtype} (Categorical): The histologic subtype proved by surgical pathology. This takes one of the following values \texttt{\{"clear\_cell\_rcc", "clear\_cell\_papillary\_rcc", "papillary",\\ "chromophobe", "urothelial", "rcc\_unclassified",\\ "multilocular\_cystic\_rcc", "wilms", "oncocytoma",\\ "angiomyolipoma", "mest", "spindle\_cell\_neoplasm"\}}.

    % CHECK
    \item \texttt{tumor\_necrosis} (Boolean): \texttt{true} if the post-operative surgical pathology report indicates that necrotic tissue is present within the tumor, \texttt{false} if the report indicates that it is not, and \texttt{null} if the report does not mention this (23 instances).
    
    % CHECK
    \item \texttt{tumor\_isup\_grade} (Integer): The WHO ISUP \cite{epstein1998world} grade of the tumor indicated in the post-operative surgical pathology report. The value of \texttt{Null} is used for cases where ISUP grade does not apply, such as benign tumors or Chromophobes.

    % CHECK
    \item \texttt{clavien\_surgical\_complications} (Categorical): This takes one of following values defined by the Clavien Dindo Grade \cite{clavien1992proposed}: \{\texttt{"0"}, \texttt{"1"}, \texttt{"2"}, \texttt{"3a"}, \texttt{"3b"}, \texttt{"4"}, \texttt{"5"}\} or \texttt{null} if this could not be determined (1 instance).
    
    % CHECK
    \item \texttt{er\_visit} (Boolean): \texttt{true} if the patient visited the ER less than 24 hours after discharge but was not admitted, \texttt{false} if not, and \texttt{null} if this could not be determined (2 instances).
    
    % CHECK
    \item \texttt{readmission} (Boolean): \texttt{true} if the patient was readmitted to a hospital within 90 days of the surgery, \texttt{false} if not, and \texttt{null} if this could not be determined (2 instances -- e.g. a censor or death date of less than 90 days after surgery).

    % CHECK
    \item \texttt{estimated\_blood\_loss} (Integer): The volume of blood in ml that the surgeon estimates was lost during the nephrectomy procedure, or \texttt{null} if this is not available (1 instance).
    
    % CHECK
    \item \texttt{surgery\_type} (Categorical): Takes one of the following values \texttt{\{"open", "laparoscopic", or "robotic"\}}.

    % CHECK
    \item \texttt{surgical\_procedure} (Categorical): Takes one of the following values \\\texttt{\{"partial\_nephrectomy", "radical\_nephrectomy"\}}.

    % CHECK
    \item \texttt{surgical\_approach} (Categorical): Takes one of the following values \\\texttt{\{"retroperitoneal", "transperitoneal"\}}.

    % CHECK
    \item \texttt{operative\_time} (Integer): The time that the nephrectomy procedure took in minutes, or \texttt{null} if this could not be retrieved (2 instances).
    
    % CHECK
    \item \texttt{cytoreductive} (Boolean): \texttt{true} if the nephrectomy was performed for debulking purposes, \texttt{false} otherwise
    
    % CHECK
    \item \texttt{positive\_resection\_margins} (Boolean): \texttt{true} if the post-operative surgical pathology report indicates that there is malignant tissue still present in the margins of the excised tissue, \texttt{false} if the report indicates that the margins are clear.

    % CHECK
    \item \texttt{last\_preop\_egfr} (Object): Information about the most recent estimated Glomular Filtration Rate (eGFR) value that was measured before the nephrectomy. In cases where no preoperative eGFR value was available, this object takes the value \texttt{null} (57 instances).
    \begin{itemize}
        \item \texttt{value} (Float): The measured value in ml/min. In cases where the value was 90 or greater, a value of \texttt{">=90"} was recorded. In cases where the patient was younger than 16 years old, GFR cannot be reliably estimated so a value of \texttt{age<16} was recorded.
        \item \texttt{days\_before\_nephrectomy} (Integer): The number of days before the nephrectomy that this measurement was taken.
    \end{itemize}
    \item \texttt{first\_postop\_egfr} (Object): Information about the first estimated Glomular Filtration Rate (eGFR) value that was measured after the nephrectomy. In cases where no postoperative eGFR value was available, this object takes the value \texttt{null} (53 instances).
    \begin{itemize}
        \item \texttt{value} (Float): The measured value in ml/min. In cases where the value was 90 or greater, a value of \texttt{">=90"} was recorded. In cases where the patient was younger than 16 years old, GFR cannot be reliably estimated so a value of \texttt{age<16} was recorded.
        \item \texttt{days\_after\_nephrectomy} (Integer): The number of days after the nephrectomy at which this measurement was taken.
    \end{itemize}
    \item \texttt{last\_postop\_egfr} (Object): Information about the most recent estimated Glomular Filtration Rate (eGFR) value that was measured after the nephrectomy. In cases where one or fewer postoperative eGFR values were available, this object takes the value \texttt{null} (122 instances).
    \begin{itemize}
        \item \texttt{value} (Float): The measured value in ml/min. In cases where the value was 90 or greater, a value of \texttt{">=90"} was recorded. In cases where the patient was younger than 16 years old, GFR cannot be reliably estimated so a value of \texttt{age<16} was recorded.
        \item \texttt{days\_after\_nephrectomy} (Integer): The number of days after the nephrectomy at which this measurement was taken.
    \end{itemize}

    % TODO vital status (verify names, look for nulls)
    \item \texttt{vital\_status} (Categorical): The current vital status of the patient. Takes one of the following values: \texttt{\{"Censored", "Dead"\}}
    \item \texttt{vital\_days\_after\_surgery} (Integer): The number of days after nephrectomy until either the censor date or the date of death.

\end{itemize}

This data has since been archived by The Cancer Imaging Archive\footnote{\url{https://wiki.cancerimagingarchive.net/pages/viewpage.action?pageId=61081171}} where the imaging and segmentations are stored in DICOM format and the clinical data has been converted to a single CSV file. Bitmaps within the JSON are flattened using two underscores, such that for example the value accessed by [``comorbidities''][``copd''] in the JSON file is stored in the CSV under the column ``comorbidities\_\_copd''.

\section{Technical Validation}
\label{technical_validation}
Any large dataset is bound to be imperfect, and this is especially true of semantic segmentation. Such datasets are still useful, of course, but their utility can be enhanced by estimating the nature and extent of these imperfections. In order to characterize the errors in our segmentation labels, we randomly selected 30 cases from the challenge's training set and repeated the image annotation process on this subset. This allowed us to estimate agreement that our annotation process has with itself, and thus assess the fidelity of the labels. We measured this agreement using the same metrics as the KiTS19 challenge. This allowed for a direct comparison to the performance of the automatic systems submitted as part of the challenge such that an automatic system that is as reliable as or better than our manual annotation process would be expected to achieve the same score as that from repeating our annotation process. The results of this study can be found in table \ref{tab:interobserver}.
\begin{table}[]
\centering
\label{tab:interobserver}
\begin{tabular}{| c | c |}
\hline
\textbf{Region}         & \textbf{Manual Mean Dice} \\
\hline
\hline
Kidney + Tumor & 0.983     \\
\hline
Tumor Only     & 0.923     \\
\hline
\end{tabular}
\vspace{1em}
\caption{The agreement of the manual annotation process with itself measured by the average S\o renson Dice score over 30 cases randomly selected from the first 210 cases.}
\end{table}

\section{Usage Notes}
\label{usage_notes}
% Data access, starter code
In addition to the release of this data, we have also released some Python starter code which includes scripts to load and visualize the data. This can be found on GitHub at \url{https://github.com/neheller/kits19}.

\section*{Acknowledgements}
\label{acknowledgements}
Research reported in this publication was supported by the National Cancer Institute of the National Institutes of Health under Award Number R01CA225435. The content is solely the responsibility of the authors and does not necessarily represent the official views of the National Institutes of Health. 

We also gratefully acknowledge Climb 4 Kidney Cancer (C4KC) for providing student scholarships which were essential to the collection and annotation of this data. C4KC is an organization dedicated to advocacy for kidney cancer patients and the advancement of kidney cancer research. More inforamtion about C4KC can be found at \url{climb4kc.org}

% TODO author contributions

% TODO competing interests

% Figures and figures legends


\bibliographystyle{splncs04}
\bibliography{main.bbl}

\end{document}